\documentclass{article}
\begin{document}

\title{Harmonic potential and hadron masses}
\author{Rafael Tumanyan, \\
Yerevan Physics Institute, Armenia} 
\maketitle
\begin{abstract}

The quark-gluon sea in the hadrons is considered as periodically correlated.  
Energy levels of Shrodinger equation with harmonic potential is used for describing 
of the spectrum of hadron masses. In the considered cases the effective potential 
operating on each particle of ensemble, under certain conditions becomes 
square-law on displacement from a equilibrium point. It can become an explanation 
of popularity of oscillator potential for the description of a spectrum of masses 
of elementary particles. The analysis shows that levels of periodic potential 
better agreed to the spectrum of hadron masses, than levels of other potentials 
used for an explanation of a spectrum of masses. 
\end {abstract}

\section{Introduction}

The description and explanation of elementary particles spectroscopy (spectrum 
of masses) is one of the important problems of modern physics. For today there 
is a number of theories and models applied to the description as interaction between quarks, and a spectrum of masses 
of the elementary particles made of quarks or partons. Phenomenological potentials in 
non relativistic case are considered in \cite{1,2,3,4}. Relativistic approach on the basis of Dirac equation 
is considered in \cite{9}. However, despite the great achievements in these areas, still there is a 
number of problems, in particular, at an explanation of a mass spectrum. The basic problem is 
obtaining of highest levels both mesons, and baryons. For an explanation of shift of masses of 
the higher D mesons found out in  \cite{10,12} it is used the dynamical QCD 
mechanism  \cite{11} in which the shift of masses calculated by introducing 
of a mixing angle of the connected channels. In the present work the new 
approach for the masses of hadrons is considered. Such 
approach is proved by  presence of partons in hadrons, which named a 
quark-gluon sea. Thus there is a natural question of the average effective field 
operating on particles of the quark-gluon sea. We will notice that in case of 
light quarks  $m_u  = 2 - 3mev,m_d  = 6 - 8mev$ \cite{12} the 
mass of a proton does not satisfy to virial theorem. If to consider a share of the neutral 
partons, presumably gluon, (50 percent of mass of a proton  \cite{5}) as potential energy 
and to add its half there are 25 percents more.   It shows that 
the proton is not simply bounded condition of 3 quarks. As well known the liquid state of 
quark-gluon matter is observed in RHIC experimental results \cite{rhic} 
and it is clear that near phase transition 
point  the quark-gluon sea density can become periodical \cite{phlett, dencor}. In this Letter we would like to 
report our study of such new state of quark-gluon sea and their effect on the hadron masses. 
Obtained results can be used for studiing of usual substances near transition point too. 

\section{Energy spectrum of periodical potential}
Under certain conditions in the quark-gluon sea can exist short range correlations, which 
means that the density and correlation functions become periodical. Therefore we can choose 
the mean field in the sea periodical too with amplitude $U_0 $
 and wave number $k = 2\pi /r_0 $

\begin{equation}
U = U_0 (1 - \varepsilon \cos kx)
\label{ucos}
\end{equation}
where the $\varepsilon $ describe the depth of correlation.
The Shrodinger equation with this potential for particle with mass $m$ and 
energy $E$ after substitution of new variable $z = kx/2$
 become a Mattieu equation  \cite{6}
\begin{equation}
\psi '' + (a - q\cos 2z)\psi  = 0
\label{eqcos}
\end{equation}
\begin{equation}
a = (E - U_0 )4m/(\hbar k)^2   \\
 \label{ownenergy}
 \end{equation}
 $$q = \varepsilon U_0 4m/(\hbar k)^2 $$
In case of small oscillation amplitude $kx \ll 1$
and we can restricted by second power of $x$ in the Taylor expansion of the potential.
This equivalent to oscillator approximation. In general case the analyse of eigenvalues of 
Mattieu equation is necessary. These eigenvalues are depend on $q$ and number of level. 
When $q$ not so big and  $r > 7$ the eigenvalues equal \cite{6}
\begin{equation}
a_n  \approx b_n +1 \approx n^2  + \frac{{q^2 }}
{{2(n^2  - 1)}} + \frac{{(5n^2  + 7)q^2 }}
{{32(n^2  - 1)(n^2  - 4)}} + ...
\label{encos}
\end{equation}
 As seen the second term  
\begin{equation}
\delta_n  = q^2 /2(n^2  - 1).
\label{addterm}
\end{equation}
 is inverse proportional to number of level and is similar to levels of 
Coulomb potential $1/r$. In certain physical cases this can explain the using 
of Coulomb potential \cite{4}.  This term is small if $q \ll 1$, but become 
considerable when $q > 1$.  Notice that such amendment can be used 
successfully for explanation of mass level 
shift in many cases of elementary particle mass. This can be realized when 
 $\varepsilon  \ll 1$. 
When  $q \gg 1$ the eigenvalues $a_n ,b_n $ number $n$
 of Mattieu functions with period $\pi $,  equal  \cite{6} ($W = 2n + 1$)
 
\begin{equation}
 a_n  \sim b_n +1 \sim  - 2q + 2W\sqrt q  - \Delta _n  
\label{nmattew}
\end{equation}

 \begin{equation}
  \Delta _n  = \frac{W}{{2^7 \sqrt q }} + \frac{{W^2  + 1}}
{8} + \frac{1}
{W}\left( {\frac{3}
{{2^7 \sqrt q }} + \frac{{33}}
{{2^{17} q^{3/2} }}} \right) 
\end{equation}
As seen the general term in this expression is $ \sim W \sim n + 1/2$
 as levels of oscillator. But the negative amendment becomes considerable for levels of 
number $ > 3$. And the higher levels becomes lower of oscillator levels 
approximately on value 
\begin{equation}
\Delta _n  \approx W^2 /8
\label{amend}
\end{equation}
The Mattieu equation is the specific type of Floque equations with periodic 
coefficients. The stable and finite solutions are solutions with $a_n  < a < b_n +1$  \cite{6}, 
which are the energy bands. The solutions with other energies are unstable and infinite.
\section{Three dimensional spectrum}
In this chapter we elucidate the influence of dimensionality on the potential 
levels. Three dimensional Shrodinger equation  have a form  
\begin{equation}	
\frac{{\hbar ^2 }}
{{2m}}\Delta \psi  + [E - U]\psi  = 0
\label{eq3d}
\end{equation}
where  $\Delta $ - is the 3-dimensional Laplace operator , $\hbar $ is the 
Plank constant. If the potential is central symmetric after substitution 
	$$
\psi (r,\theta ,\phi ) = R(r)Y_{lm} (\theta ,\phi )
$$

one can find
\begin{equation}	
\frac{1}
{{r^2 }}\frac{d}
{{dr}}\left( {r^2 \frac{{dR}}
{{dr}}} \right) - \frac{{l(l + 1)}}
{{r^2 }}R + \frac{{2m}}
{{\hbar ^2 }}[E - U(r)]R = 0
\label{radeq}
\end{equation}
here $Y_{lm} $ - are the eigenfunctions with azimuth number $l$
 and magnetic number $m$. As seen from additional second term in our 
 approach the energy  levels are degenerated. But consideration of 
 3-dimensionality to take off the degeneracy and split the levels. The ground 
state $l = 0$
 is not degenerate and is equivalent to one dimensional case. The case  $l > 0$
 we consider qualitative in the approximation $r = r_{eff}$ in the potential. 
 In this approximation appears additional term in the expression of energy 
$\delta E_l  = \hbar ^2 l(l + 1)/2mr_{eff}^2 $, which split the levels with 
$l \ne 0$. Calculation by perturbation theory give amendment with 
$r_{eff}  = r_n $, where $r_n $
 is the mean square radius of the level number $n$. It is not difficult to show 
 that splitting is equidistant in coincide with experimental data. It is known \cite{5}, 
 that the highest level is splitted to three levels. So this is 
the fourth level with $l = 0,1,2$. 
The whole classification of hadrons by using of oscillator levels can be used 
in our case too.
\section{Application to mass spectrum}
Calculations show that spectrum ($ \sim n^2 $) can applicable to both mesons 
and baryons. The meson masses are raise approximately by law 
\begin{equation}	
m_n  = m_1  + m_2 n^2 
\label{massform}
  \end{equation}
In this sequence are included long life mesons with $\pi $
 in the beginning {$\pi, K and  \eta, \Phi, D, \psi, \psi' $}. The mass agreement
 is good if one choose  $U_0  = m_1  \approx 15 Mev$, but multiplier of $n^2 $
 approximately $125 Mev$.  The calculation of characteristic length parameter 
 $r_0 $ gives a value $r_0  = 0.25 Fm $
 if we choose the quark mass $m_q  = 2Mev$ \cite {12}. In case of higher mass 
 of the quark this is decreased more. But the charge radius of 
mesons is about 0.67 Fm \cite{12}.  But we must remember, that meson 
charge radius is scattering length, but is not the particle 
real physical radius. The comparison of high levels of mesons, for example 
$\psi $, as quark-anti quark system, shows that high levels do not agree to 
oscillator levels \cite{5}. Possible, they agree with (\ref{nmattew}), where the 
condition $q \gg 1$ is satisfied and levels have shift down  \cite{5}. The relative 
 amendment $\delta _n  = \Delta _n /2W\sqrt q  = W/16\sqrt q $ of levels $n = 4 - 6$
 is about 5-10 percents, if $\sqrt q  = 5 - 10$, 
 which give a good agreement of levels to real masses of particles. 
Calculation shows, that this possible, if the characteristic length two times longer. 
Possible, this is a second harmonic of some periodic potential. This spectrum 
is better agreed to real mass spectrum than an energy spectrum of any 
other potential. Notice that obtained levels $a_n ,b_n +1$ are very close each 
other, which means possible the degeneration on spin due to quarks are fermions. 
Mass levels of the same form (\ref{massform}) describe baryon masses 
if we add to nucleon $N$ mass this expression with $m_1  = 155Mev$, and 
$m_2  = 25Mev$. This row include {$N, \Lambda, \Sigma, \Theta, \Omega$}.
The characteristic length is about 0.85 Fm, which is close to the 
charge radius of proton  \cite{12}. So the characteristic lengths are close to 
particles characteristic radii. But we must remember, that these radii are 
scattering lengths, but these are not the particle real physical 
radii. Taking into account the relation for energy 
\begin{equation}	
E_n  = \frac{{(\hbar k)^2 }}
{{4m}}a_n  + U_0 
\label{enrel}
\end{equation}
it is clear that mesons and baryons mass formulae differs in $1/mr_0^2 $
 and $U_0 $.
In our approach the agreement of calculated masses to real hadron 
masses considerably improves. Thus, the assumption of periodicity of 
potential removes set of problems and gives 
more adequate picture of potential inside hadrons. Notice, that the second 
term \ref{addterm} ib the \ref{encos} in case $q=1-4$ can be used for obtaining of 
low bound states and energy levels similar to very strong Coulomb force.
But the rich and complicated spectrum of hadron masses, most likely, means presence of 
periodic potential with several Fourier components. This explains also taking place well-
known violation of strong interaction by middle strong interaction which is 
approximately ten times weaker than the strong. And, well-known that growth 
of mass of particles on linear law relative to  level number in some cases is 
observed. Only the periodic potential depending on parameters of a problem 
has the spectrum creating a basis for the description of all such cases. Let's 
notice that at $q \gg 1$ eigenvalues of energy of periodic potential more coincide 
to masses of systems a quark-anti quark, including the higher levels which 
is not possible to do by any other method. All other methods are introducing  
new parameters for obtaining of levels agreement, for example mixing angle et al. 
\section{Summary}
Studying of a energy spectrum of the equation of Shrodinger with harmonic 
potential shows that it suitable for an explanation of a spectrum of masses 
more than other potentials used. The 
spectrum of harmonic potential with small amplitude give sequence of masses 
of meson and baryon nonets. The additional term is inverse proportional to 
square of level number and is similar to levels of Coulomb potential. The 
spectrum of harmonic potential with great amplitude is similar to a oscillator 
spectrum, but there is a negative amendment to levels which grows quadratic 
on number of level and reduces high levels. Such spectrum explains also high 
levels of mesons, considered as pair a quark-anti quark, charmonium and etc.
The account of 3 dimensionality leads to splitting of levels by removal of degeneration on 
the orbital moment. Thus, the assumption of 
periodicity of an effective field inside hadrons due to periodic spatial distribution 
of partons gives good agreement with a spectrum of hadron masses .  But
for a complete description, probably, it is necessary to consider few harmonics. Such 
approach, most likely, will create possibility of a uniform method to describe the whole 
spectrum of  hadron masses as it have the energy levels both quadratic and linear 
on level number in dependence of task parameters. Notice that we use only 
two mass parameters instead of two parameters ($\alpha _s ,k$) used by others. 
Beside this the high levels are obtained too. Any other method uses new 
parameters, for example, mixing angle, for obtaining higher levels. It is very 
surprising that the periodical potential exist in the hadrons. But as known the 
liquid state of quark-gluon sea is observed at RHIC and under certain conditions 
constituents of this sea can arranged periodiacally near the phase transition point. 
Our results can be used for better understanding of RHIC results and hadron structure.

\end{document}